\documentclass[1p]{elsarticle}
\usepackage{amsmath,amssymb,amsfonts}
\usepackage{graphics}
\usepackage{epsfig}

\newcommand\al{\alpha}

\newcommand\be{\beta}

\newcommand\ga{\gamma}

\newcommand\De{\Delta}
\newcommand\Ga{\Gamma}

\newcommand\lam{\lambda}

\newcommand\ce{{\cal E}}

\newcommand\ep{\epsilon}

\newcommand\MD{\mathfrak{D}}
\newcommand\BMD{\bar{\mathfrak{D}}}

\begin{document}
\begin{frontmatter}
\title{Fast GPU-based calculations in few-body quantum scattering}
\author{V.N. Pomerantsev}
\ead{pomeran@nucl-th.sinp.msu.ru}
\author{V.I. Kukulin}
\ead{kukulin@nucl-th.sinp.msu.ru}
\author{O.A. Rubtsova}
\ead{rubtsova@nucl-th.sinp.msu.ru}
 \address{Skobeltsyn Institute of
 Nuclear Physics, Lomonosov Moscow State University, Leninskie gory 1(2), Moscow,
119991, Russia}
\author{S.K. Sakhiev}
\ead{ssayabek@yandex.kz}
\address{L.N. Gumilyov Eurasian National University,
Astana, 010000, Kazakhstan}

\begin{abstract}
A principally novel approach towards solving the few-particle
(many-dimensional) quantum scattering problems is described. The
approach is based on a complete discretization of few-particle
continuum and usage of massively parallel
computations of  integral kernels for scattering equations by
means of GPU. The  discretization for
continuous spectrum of a few-particle Hamiltonian is realized with
a projection of all scattering operators and wave functions onto
the stationary wave-packet basis. Such projection procedure leads
to a replacement of singular multidimensional integral equations
with linear matrix ones having finite matrix elements. Different
aspects of the employment of a multithread GPU computing for  fast
calculation of the matrix kernel of the equation are studied in
detail. As a result, the fully realistic three-body scattering
problem above the break-up threshold   is solved on an ordinary
desktop  PC with GPU for a rather small computational time.

\end{abstract}

 \end{frontmatter}
\section{Introduction}

Solution of few-body scattering problems, especially above the
three-body breakup threshold, no matter in differential or integral
formalism, involves a very large amount of calculations and
therefore requires extensive use of modern computational facilities
such as powerful supercomputers. As a vivid  example, we note that
one of the most active and successful groups in the world in this
area --- the Bochum--Cracow group guided up to recent time by Prof.
W.~Gl{\"o}ckle (who passed away recently) --- employed for such
few-nucleon calculations the fastest in Europe supercomputer from
JSC in J{\"ulich} with the architecture of Blue Gene
\cite{gloeckle12,gloeckle}. Quite recently, new methods for solving
Faddeev and Faddeev-Yakubovsky few-body scattering equations using
(in one way or another) the bases of square-integrable functions
have been developed \cite{Lazauskas_rep}, which allow to simplify
significantly the numerical solution schemes. Nevertheless, the
treatment of realistic three- and four-body scattering problems
includes a tremendous numerical labor and, as a result, still can be
done only by a few groups over the world that hinders the
development of these important studies.

However, recently there appeared a new possibility to use the Graphics
Processing Units (GPU) for such time-consuming calculations. This
can transform an ordinary PC into a supercomputer. There is no necessity to argue that
such variant is unmeasurably cheaper and more accessible for many
researchers in the world.  However, due to the special GPU
architecture usage of GPU is effective only for those problems where
numerical schemes of solution  can be realized with a high degree
of parallelism. The high effectiveness of the so-called General
Purpose Graphics Processing Unit (GPGPU) computing has been
demonstrated in many areas of quantum chemistry, molecular
dynamics, seismology, etc. (see the detailed description of
different GPU applications in refs.
\cite{CUDA,CUDA_MC,CUDA_QCD,CUDA_ch}). Nevertheless, according to
present authors' knowledge, GPU computing still has not been used
widely for a solution of few-body scattering problems (we know
only two researches  but they are dedicated  to the {\em ab
initio}  calculation of bound states \cite{Vary} and also
resonances in the Faddeev-type formalism \cite{yarevsky}). Thus,
in this paper we would like to study in detail just the
effectiveness of GPU computing in solving general few-body
scattering problems.

In the case when the colliding  particles have inner structures and
can be excited in the scattering process, i.e. should be treated as
composite ones (e.g.,  nucleon isobars) the numerical complexity of
the problem is increased additionally, so that without a significant
improvement of the whole numerical scheme the practical solution
of such multichannel problems becomes to be highly  nontrivial
even for a supercomputer. Thus, the development of new methods in
few-body scattering which can be adapted for massively parallel
realization is of interest nowadays. We propose here a novel
approach in this area which includes two main components:

(i) A complete discretization of the continuous spectrum of the
scattering problem, i.e. the replacement of continuous momenta and
energies  with their discrete counterparts, by projecting all the
scattering functions and operators onto a space spanned on the
basis of the stationary wave packets
\cite{KPR_Ann,KPRF_breakup,Kelvitch_Yaf14,KPR_GPU}. As a result,
the integral equations of the scattering  theory (like the
Lippmann--Schwinger, Faddeev etc. equations) are replaced with
their matrix analogs. Moreover, due to an ordinary $L_2$
normalization of the stationary wave packets one can solve a
scattering problem almost fully similarly to  bound-state
problems, i.e. without explicit account of the boundary conditions
(which are rather nontrivial above few-body breakup thresholds).
The main feature of this discretization procedure is that all the
constituents in the equations are represented with finite
matrices, elements of which are calculated independently. So, this
approach is just quite suitable for parallelization and
implementation on GPU.

(ii) The numerical solution of the resulting matrix equations with
wide usage of the multithread computing on GPU.

In the present paper, we adapt the general wave-packet
discretization algorithm for  GPU implementation by an example of
calculating the elastic scattering amplitude in three-nucleon
system with realistic interactions. Also different aspects related
to GPU computing are studied and runtimes for CPU and GPU mode
calculations are compared.

The paper is organized as follows. In the section II we briefly
recall the main features of the wave-packet continuum
discretization approach towards solving two- and three-body
scattering problems. The  numerical scheme for a practical
solution of the $nd$ elastic scattering problem in a discretized
representation is described in Section III. In the next Section IV
we discuss the properties of GPU computing for the above problem
and test some illustrative examples while in the Section V the
results for the $nd$ elastic scattering with realistic $NN$
interaction are presented. The conclusions are given in  the last
Section VI.

\section{Continuum discretization with stationary wave-packets  in few-body
scattering problems}

In this section we outline briefly the method of stationary wave
packets that is necessary for understanding the subsequent material
by reader. For detail we refer to our previous original papers
\cite{KPRF_breakup,KPR_GPU} and the recent review \cite{KPR_Ann}.

\subsection{Stationary wave packets for two-body Hamiltonian}
%Stationary wave-packets have a meaning of normalized continuum
%states for some Hamiltonian.
Let us introduce some two-body Hamiltonian $h=h_0+v$ where $h_0$
is a free Hamiltonian (the kinetic energy operator) and $v$ is an
interaction.  Stationary wave packets (WPs) are constructed as
integrals of exact scattering wave functions
$|\psi_p\rangle$ (non-normalized) over some
 momentum intervals
$\{\De_i\equiv[p_{i-1},p_i]\}_{i=1}^N$:
\begin{equation}
\label{z_k}
|z_k\rangle=\frac{1}{\sqrt{C_k}}\int_{{\De}_k}w(p)|\psi_p\rangle dp,
\quad C_k=\int_{\De_k}|w(p)|^2dp.
\end{equation}
Here $p=\sqrt{2mE}$ is relative momenta, $m$ is the reduced mass of
the system, $w(p)$ is a weight function and $C_k$ is the
corresponding normalization factor.

The set of WP states (\ref{z_k}) has a number of interesting and
useful properties \cite{KPR_Ann}. First of all, due to the
integration in eq.~(\ref{z_k}), the WP states have a finite
normalization as bound states. The set of WP functions together with
the possible bound-state wave functions $|z^b_n\rangle$ of the
Hamiltonian $h$ form an orthonormal set and can be employed as a
basis  similarly to any other $L_2$ basis functions, which are used
to project wave functions and operators \cite{KPR_Ann}. (To simplify
notations, we will omit below the superscript $b$ for bound-states
and will differ them from WP states just by their index $n\leq
N_b$.)

The matrix of Hamiltonian $h$  is diagonal in such a WP basis. The
resolvent $g(E)=[E+{\rm i}0-h]^{-1}$ for Hamiltonian $h$ has also
a diagonal representation in  the subspace spanned on the WP
basis:
\begin{equation}
\label{g_tot} g(E)\approx \sum_{n=1}^{N_b}\frac{|z_n\rangle
\langle z_n|}{E-\ep_n^*}+\sum_{k=N_b+1}^N|z_k\rangle g_k(E)
\langle z_k|,
\end{equation}
where $\ep_n^*$ are the bound-state energies and  eigenvalues
$g_k(E)$ can be expressed by explicit formulas \cite{KPR_Ann}.

\subsection{Free wave-packet basis}
Useful particular case of stationary wave packets is free WP
states which are defined for the free Hamiltonian  $h_0$.
 As in the general case, the  continuum of $h_0$
  (in every spin-angular channel  $\alpha$) is divided
   onto non-overlapping intervals
 $\{\mathfrak{D}_i\equiv[\ce_{i-1},\ce_i]\}_{i=1}^N$
and  two-body free wave-packets are introduced as integrals of exact
free-motion wave functions $|p\rangle$ (an index $\alpha$ which marks
possible quantum numbers we will omit where is possible):
\begin{equation} \label{ip}
|x_i\rangle=\frac{1}{\sqrt{B_i}}\int_{\mathfrak{D}_i}f(p)|p\rangle
dp,\quad B_i=\int_{\MD_i}|f(p)|^2 dp,
\end{equation}
where $B_i$ and $f(p)$  are the normalization
factor and weight function respectively.

As has been mentioned above, in such a basis,  the free
Hamiltonian $h_0$ has a diagonal finite-dimensional representation as well as the free  resolvent
$g_0=[E+{\rm i}0-h_0]^{-1}$:
\begin{equation}
{g}_0(E)\approx\sum_{i=1}^N|x_i\rangle g_{0i}(E) \langle x_i|,
\end{equation}
where eigenvalues $g_{0i}(E)$ have analytical expressions \cite{KPR_Ann}.

Besides the above useful properties which are valid for any wave
packets, the free WP states have some other important features. In
momentum representation, the  states  (\ref{ip}) take the form of
step-like functions:
\begin{equation}
\label{theta_p}
\langle p|x_i\rangle=\frac{f(p)\theta(p\in \MD_i)}{\sqrt{B_i}},
\end{equation}
where the Heavyside-type theta-function is defined by the conditions:
\begin{equation}
\theta(p\in\MD_i)=\left\{
\begin{array}{cr}
1,&p\in\MD_i,\\
0,&p\notin\MD_i.\\
\end{array}
\right.
\end{equation}
In practical calculations, we usually used the free WP states with unit weights
$f(q)=1$.
The functions of such states  are constant inside momentum intervals.
In few-body and multidimensional cases, the WP bases are constructed as direct
products of two-body ones, so that the model space can be considered as
a multidimensional lattice.

Thus, the explicit form of the free WPs makes them very convenient  for use as
a basis in the scattering calculations \cite{KPR_Ann}. For
example, the special form of the basis functions in the momentum
representation allows to find easily the matrix elements of the
interaction potential in the free WP representation using the
original momentum representation $v(p,p')$ for the potential:
\begin{equation}
\label{vpot}
v_{ii'}=\frac1{\sqrt{B_iB_{i'}}}
\int_{\MD_i}\int_{\MD_{i'}}dpdp'f^*(p)v(p,p')f(p').
\end{equation}
Moreover, in some rough approximation the potential matrix elements
can be found simply as
$v_{i,i'}\approx\sqrt{B_iB_{i'}}v(p_i^*,p_{i'}^*)$,
where $p_i^*$ and $p_{i'}^*$ are the middle values of momenta in the intervals
${\MD_i}$ and ${\MD_{i'}}$ respectively.
Further we will use the above free WP representation for solution of
scattering problems.

It was shown \cite{KPR_Ann}, that the scattering WPs (\ref{z_k})
for some total Hamiltonian $h$ can be also approximated in the
free WP representation. There is no necessity to find the exact
scattering wave functions $|\psi_p\rangle$ in that case. Instead,
it is just sufficient to diagonalize the total Hamiltonian matrix
in the basis of free WPs. As a result of such direct
diagonalization one gets the approximate scattering WPs (and also
the functions of bound states if they exist) for Hamiltonian $h$
in the form of expansion into free WP basis:
\begin{equation}
\label{exp_z}
|z_k\rangle\approx \sum_{i=1}^N O_{ki}|x_i\rangle,
\end{equation}
where $O_{ki}$ are the matrix elements for rotation from one basis
to another. Note that it is not required that the potential $v$ is a
short-range one. So that, the same procedure allows to construct
wave packets for Hamiltonian including the long-range Coulomb
interaction and to get an analytical finite-dimensional
representation for the Coulomb resolvent~\cite{KPR_Ann}.

%%%%%%%%%%%%%%%%%%%%%%%%%%
\subsection{Scheme for a solution of a two-body scattering problem}
%%%%%%%%%%%%%%%%%%%%%%%%%%%%%%
Let us briefly discuss how to solve a two-body scattering problem
in a free WP basis. The Lippmann--Schwinger equation for the
transition operator $t(E)$
\begin{equation}
\label{lse_op} t(E)=v+vg_0(E)t(E),
\end{equation}
where $g_0(E)$ is the free resolvent,
  has
the following form in  momentum representation (e.g. for every
partial wave $l$):
\begin{equation}
\label{lse} t_l(p,p';E)=v_l(p,p')+\frac1{4\pi}\int_0^\infty dp''
\frac{v_l(p,p'')t_l(p'',p';E)}{E+{\rm i}0-\frac{(p'')^2}{2m}}.
\end{equation}
 By
projecting the eq.~(\ref{lse_op}) onto the free WP basis, the
integral equation is reduced to a matrix one in which all the
operators are replaced with their  matrices in the given basis. In
the resulting equation the momentum variables are discrete but the
energy variable remains continuous. So, in order to get the
completely discrete representation one can employ some additional
energy averaging for a projection of the free resolvent. In WP
representation, this means an averaging of its eigenvalues
${g}_{0i}(E)$:
\begin{equation}
{g}_{0i}(E)\to [{g}_0]_i^k=\frac1{D_k}\int_{\MD_k}dE\,{g}_{0i}(E),\quad
E\in\MD_k
\end{equation}
where $D_k=\ce_k-\ce_{k-1}$ is the width of the on-shell energy interval.

As a result, the WP analog for the transition operator can be
found
 from solution of the matrix equation in the free WP
representation :
\begin{equation}
\label{lsk} t^k_{ii'}=v_{ii'}+ \sum_{j=1}^N v_{ij}[{g_0}]^k_j
t^k_{ji'},\quad E\in\MD_k
\end{equation}
where $v_{ij}$ are the matrix elements of the interaction operator
which are defined by the eq.~(\ref{vpot}).
  Then the solution of the eq.~(\ref{lsk}) takes the form of histogram
  representation for the off-shell $t$-matrix from eq.~(\ref{lse})
\begin{equation}
t_l(p,p';E)\approx \frac{t_{ii'}^k}{\sqrt{D_iD_{i'}}},\quad
\begin{array}{c}
p\in\MD_i,\\
p'\in\MD_{i'},\\
E\in\MD_k,
\end{array}
\end{equation}
where $D_i$ and $D_{i'}$ are the widths of energy intervals.

As is clear, the above WP approach has some similarities to the
methods which somehow employ a discrete momentum representation,
such as a direct solution of the integral equation (\ref{lse}) with
using mesh-points or the lattice method. However, the main
difference from those is that, in addition to introducing
mesh-points for a discretization, we average the kernel functions on
momentum and energy by an integration within energy intervals (or
over the lattice cells in a few-body case). In this way, all
possible singularities in the integral kernels are somehow smoothed
out, and instead the continuous momentum dependence one has  finite
regular matrices for all operators. Moreover, all intermediate
integrations in the integral kernel can be easily performed with
using the WP projection, so that each operator in such a product is
represented as a separate matrix in the WP representation.

All  these features render the solution of scattering problems
quite similar to that for a bound-state problem (e.g. with matrix
equations and without an explicit matching with boundary
conditions etc). Besides that, this fully discrete matrix form for
all scattering equations is very suitable for parallelization and
multithread implementation (e.g. on GPU).

\subsection{Three-body wave-packet basis}
The method of continuum discretization described above is directly
generalized to the case of three- and few-body system. For a general
three-body scattering problem it
is necessary to define  WP bases for each  set of Jacobi momenta $(p_a,q_a)$, ($a=1,2,3$).
Below we show how to define the free and the channel three-body WP
states for one  Jacobi set corresponding to the $\{23\}1$
partition of the three-body system.

%\subsubsection{The lattice wave-packet basis}
For the given Jacobi partition $\{23\}1$, it is appropriate
to consider three two-body subHamiltonians: the free subHamiltonian
$h_0$ corresponding the free motion over relative momentum $p$
between particles 2 and 3; the subHamiltonian $h_1=h_0+v_1$
 which includes an interaction $v_1$ in the subsystem $\{23\}$ and also
 the free subHamiltonian $h_0^1$ corresponding to the free motion
 of the spectator particle 1 (over momentum $q$). These
 subHamiltonians form two basic three-body Hamiltonians
\begin{equation}
\label{chan} H_0=h_0\oplus h_0^1,\quad H_1=h_1\oplus h_0^1,
\end{equation}
where $H_0$ is a three-body free Hamiltonian while the channel
Hamiltonian $H_1$ defines
three-body asymptotic states for the partition $\{23\}1$. The WP
approach allows to construct basis states for both Hamiltonians
$H_0$ and $H_1$.

  At first we define the
three-body free WP basis by introducing partitions of the continua
for two free subHamiltonians $h_0$ and $h_0^1$ onto non-overlapping
intervals $\{\mathfrak{D}_i\equiv[\ce_{i-1},\ce_i]\}_{i=1}^N$ and
$\{\bar{\mathfrak{D}}_j\equiv[\bar\ce_{j-1},\bar\ce_j]\}_{j=1}^{\bar
N}$  and  two-body free WPs as in eq.~(\ref{ip}) respectively. Here
and below we denote functions and values corresponding to the
 $q$ with additional bar mark to distinguish them from the
functions corresponding to the momentum $p$.

The three-body free
WP states are built as direct products of the respective two-body
WP states. Also one should take
into account  spin and angular parts of the basis functions. Thus
the three-body basis functions can be written as:
\begin{equation}
\label{xij} |X_{ij}^{\Ga\al\be}\rangle\equiv |x_i^\al,{\bar
x}_j^\be;\al,\be:\Ga\rangle=|x_i^\al\rangle\otimes|{\bar
x}_j^\be\rangle |\alpha,\beta:\Gamma\rangle,
\end{equation}
where $|\alpha\rangle$ is a spin-angular state for the $\{23\}$
pair, $|\beta\rangle$  is a  spin-angular state of the third
particle, and $|\Gamma\rangle$ is a set of  three-body quantum
numbers. The state (\ref{xij}) is a WP analog of the exact plane
wave state in three-body continuum $|p,q;\alpha,\beta:\Gamma\rangle$
for the three-body free Hamiltonian $H_0$.

The  three-body free WP basis functions (\ref{xij}) are constant
inside the rectangular cells of the momentum lattice built from two
one-dimensional cells $\{\mathfrak{D}_i\}_{i=1}^{N}$ and
$\{\bar{\mathfrak{D}}_j\}_{j=1}^{\bar N}$ in momentum space. We
refer to the free WP basis as {\em a lattice} basis and denote
the respective two-dimensional bins (i.e. the lattice cells) by
$\MD_{ij}=\MD_i\otimes\BMD_j$. Using such a basis one can construct
finite-dimensional (discrete) analogs of the basic scattering operators.

To construct the WP basis for the channel Hamiltonian
(\ref{chan}), one has to introduce scattering wave-packets
corresponding to the subHamiltonian $h_1$ according to
eq.~(\ref{z_k}).
 The states (\ref{z_k}) are orthogonal to bound-state wave functions
and jointly with the latter they form a basis for the
subHamiltonian $h_1$. To construct  these states we employ
here a diagonalization procedure for $h_1$ subHamiltonian matrix in
the free WP basis and further one uses  the expansion
(\ref{exp_z}).

  Now the  three-body wave-packets for the
  channel Hamiltonian $H_1$ are defined
just as products of two types of wave-packet states for $h_1$ and
$h_0^1$ subHamiltonians
 whose  spin-angular parts are combined to
the respective three-body states having quantum numbers $\Ga$:
 \begin{equation}
\label{si} |Z^{\Ga\al\be}_{kj}\rangle
\equiv|z_k^\al,\bar{x}_j^\be,\al,\be:\Ga\rangle,\quad
\begin{array}{c}{k=1,\ldots,N},\\ j=1,\ldots,\bar{N}.
\end{array}
\end{equation}

 The properties of such WP states (as well as the properties of free WP states)
have been studied in detail in a series of our previous papers
(see e.g. the review \cite{KPR_Ann} and references therein to the
earlier works). In particular, they form an orthonormal set and
any three-body operator which functionally depends on the channel
Hamiltonian $H_1$ has a diagonal matrix representation in the
subspace spanned on this basis. It allows us to construct an
analytical finite-dimensional approximation for the {\em three-body
channel resolvent} $G_1(E)\equiv [E+{\rm i}0-H_1]^{-1}$ which
enters the Faddeev-equation kernel \cite{KPR_Ann,KPRF_breakup}.

The simple analytical representation  for the channel three-body
resolvent $G_1(E)$ is one of the main features for the wave-packet
approach since it allows to simplify enormously the whole
calculation of  integral kernels and thereby to simplify solving general
three- and few-body scattering problems.

\section{Discrete analogue for Faddeev equation for 3N system
 in the wave-packet representation}
We will illustrate a general approach to solving few-body scattering
problems by an example of scattering  in a system of three identical
particles using the Faddeev framework, namely the
elastic $nd$ scattering (treatment of the three-body breakup in $3N$
system has been discussed in ref.~\cite{KPRF_breakup}). In this
case, the system of Faddeev equations for  the transition operators
(or for the total wave function components) is reduced to a single
equation. So that, the WP basis is defined for one Jacobi coordinate
set only.

\subsection{The Faddeev equation for a transition operator}
%%%%%%%%%%%%%%%%%%%%%%%%%%%%%%%%%%%%%%%%%%%%%%%%%%%%%%%%%%
The elastic scattering observables can be found from the single
Faddeev equation  for the transition operator $U$,
 e.g. in the following form (the so-called Alt-Grassberger-Sandhas form):
\begin{equation}
\label{pvg} U=Pv_1+Pv_1G_1U.
\end{equation}
Here $v_1$ is the pairwise interaction between particles 2 and 3,
$G_1$ is the resolvent of the channel Hamiltonian $H_1$, and
$P$ is the  particle permutation operator defined as
\begin{equation}
P=P_{12}P_{23}+P_{13}P_{23}.
\end{equation}
Note that the operators of this type enter the kernels of the
Faddeev-like equations in general case of  non-identical particles as well.
So that, the presence of the permutation operator $P$ is a
peculiar feature of the Faddeev-type kernel which causes major
difficulties in a practical solution of such few-body scattering
equations.

After the partial wave expansion in terms of  spin-angular
functions, the operator equation (\ref{pvg}) for each value of the
total angular momentum and parity is reduced to a system of
two-dimensional singular integral equations in momentum space. The
practical solution of this system of equations is complicated and
time-consuming task due to special features of the integral kernel
and a large number of coupled spin-angular channels which should be
taken into account \cite{gloeckle}.

In particular, the Faddeev kernel at the real total energy has
singularities of two types: two-particle cuts corresponding to
all bound states in the two-body subsystems and the three-body
logarithmic singularity (at energies above the breakup threshold).
While the regularization of the two-body singularities is
straightforward and does not pose any problems, the regularization of
 the three-body singularity requires some special
techniques that greatly hampers the solution procedure. The
practical tricks which allows to avoid such complications are e.g. a
solution of the equation at complex values of energy followed by
analytic continuation to the real axis or a shift for the contour of
integration from the real axis in the plane of complex momenta.

However, the main specific feature of the Faddeev-like kernel is the
presence of the particle permutation operator $P$, which changes the
momentum variables from one Jacobi set to another one. Integral
kernel of this operator $P(p,q;p',q')$ as a function of the momenta
contains the Dirac $\delta$-function and two Heaviside
$\theta$-functions \cite{gloeckle}, so the double integrals in the
integral term have variable limits of integration. Therefore, when
replacing the integrals with the quadrature sums it is necessary  to
use a very numerous multi-dimensional interpolations of the unknown
solution from a ``rotated'' momentum grid to the initial one. This
cumbersome interpolation procedure takes most of the computational
time and requires using powerful supercomputers.

 The WP discretization
method described here allows to circumvent completely the above
difficulties in solving the Faddeev  equations (see \cite{KPR_Ann}
and below).

 \subsection{The matrix analog of the Faddeev equation and its features}

As a result of projecting  the integral equation (\ref{pvg}) onto
the three-body  channel WP basis (\ref{si}), one gets its matrix
analog (for each set of three-body quantum numbers $\Gamma$):
\begin{equation}
\label{m_pvg} {\mathbb U}={\mathbb P}{\mathbb V}_1+{\mathbb
P}{\mathbb V}_1 {\mathbb G}_1 {\mathbb U}.
\end{equation}
Here ${\mathbb P}$, ${\mathbb V}_1$ and ${\mathbb G}_1$ are the
matrices of the permutation operator, pair interaction and channel
resolvent respectively defined in the channel WP basis.

While the matrices of the pairwise interaction and channel
resolvent ${\mathbb V}_1$ and ${\mathbb G}_1$ in WP basis can be
easily evaluated \cite{KPR_Ann,KPR_GPU}, the calculation of
 permutation matrix ${\mathbb P}$  is not quite trivial task.

However the permutation operator matrix $\mathbb P$ in the
three-body channel WP basis  can be expressed through the
 matrix ${\mathbb P}^0$ of the same operator   in the lattice basis (\ref{xij}) using the
rotation matrices $\mathbb O$ from the expansion (\ref{exp_z})
(which depend on spin-angular two-particle state $\alpha$):
 \begin{eqnarray}
  \label{perm_z}
 [\mathbb{P}]^{\Ga\al\be,\al'\be'}_{kj,k'j'}
  \approx
\sum_{ii'}O_{ki}^\alpha O_{k'i'}^{*\alpha'}
[\mathbb{P}^0]^{\Ga\al\be,\al'\be'}_{ij,i'j'},\qquad\qquad\qquad\qquad\\
\nonumber [\mathbb{P}]^{\Ga\al\be,\al'\be'}_{kj,k'j'}\equiv \langle
Z_{kj}^{\Ga\alpha\beta}|P|Z_{k'j'}^{\Ga\alpha'\beta'}\rangle,\quad
[\mathbb{P}^0]^{\Ga\al\be,\al'\be'}_{ij,i'j'}\equiv \langle
X_{ij}^{\Ga\alpha\beta}|P|X_{i'j'}^{\Ga\alpha'\beta'}\rangle.
\end{eqnarray}

A matrix element of the operator $P$ in the lattice basis is
proportional to the overlap between basis functions defined in
different Jacobi sets \cite{KPRF_breakup}.
 Such a matrix element can be
calculated by integration with the weight functions over the momentum
lattice cells:
\begin{eqnarray}
[\mathbb{P}^0]^{\Ga\al\be,\al'\be'}_{ij,i'j'} =
\int_{\MD_{ij}}p^2dpq^2dq\int_{\MD'_{i'j'}}(p')^2dp'(q')^2dq'\times\nonumber\\
\frac{f^*(p)\bar{f}^*(q)f(p')\bar{f}(q')}{\sqrt{B_iB_{i'}\bar{B}_j\bar{B}_{j'}}}\langle
pq,\al\be:\Ga|P|p'q',\al'\be':\Ga\rangle,\label{perm}\qquad \qquad
\end{eqnarray}
where the prime at the lattice cell   $\MD'_{i'j'}$ indicates that
the cell belongs to the rotated Jacobi set while $\langle
pq,\al\be:\Ga|P|p'q',\al'\be':\Ga\rangle$ is the kernel of particle
permutation operator in a momentum space which can be written in the
 form:
\begin{equation}
\label{gal} \langle pq,\al\be:\Ga|P|p'q',\al'\be':\Ga\rangle=
\sum_{\ga\ga'}g^{\Ga\al\be,\al'\be'}_{\ga\ga'}I_{\ga\ga'}(p,q,p',q'),
\end{equation}

where $\ga$ and $\ga'$ represents the intermediate three-body spin-angular
quantum numbers, $g_{\ga\ga'}$ are algebraic coupling coefficients and the
function $I_{\ga\ga'}(p,q,p',q')$ is proportional to the product of the Dirac
delta and Heaviside theta functions \cite{gloeckle}. However, due to integration
in the eq.~(\ref{perm}), corresponding energy and momentum singularities get
averaged over the cells of the momentum lattice and, as a result, the elements
of  the permutation operator matrix in the WP basis are finite. Finally, the
matrix element (\ref{perm})  is reduced to a double integral with variable
limits and can be  calculated  numerically \cite{KPR_GPU}.

The on-shell elastic amplitude for the $nd$ scattering in the WP
representation is defined now via
 the diagonal  matrix element of the $\mathbb
U$-matrix \cite{KPR_Ann}:
\begin{equation}
A_{\rm el}^{\Gamma\al_0\be}(q_0)\approx  \frac{2m}{3q_0}
\frac{[{\mathbb{U}}]^{\Ga\al_0\be,\al_0\be}_{1j_0,1j_0}}
{\bar{d}_{j_0}},
\end{equation}
 where $m$ is the nucleon  mass, $q_0$ is the initial two-body momentum
 and the matrix element is taken between the channel WP states $|Z^{\Ga\alpha_0\beta}_{1j_0}\rangle=|z_{1}^{\alpha_0},{\bar
 x}_{j_0}^\lam;\al_0,\be:\Ga\rangle$
corresponding to the initial and final scattering states. Here
$|z_{1}^{\al_0}\rangle$ is the bound state of the  pair, the index
$j_0$ denotes the bin $\BMD_{j_0}$ including the on-shell momentum
$q_0$ and $\bar{d}_{j_0}$ is a momentum width of this bin.

It should be noted here that,  in our discrete WP approach,   {\em
the three-body breakup is treated as a particular case of
inelastic scattering} \cite{KPRF_breakup} (defined by the
transitions to the specific two-body discretized continuum
states), so that the breakup amplitude  can be found in terms of
{\em the same matrix} $\mathbb U$ determined from the
eq.~(\ref{m_pvg}). This feature gives an additional  advantage to
the present WP approach.

\subsection{\label{secIIIc} The features of the numerical scheme for solution in WP approach}
%%%%%%%%%%%%%%%%%%%%%%%%%%%%%%%%%%%%%%%
So, in the WP approach,  we reduced the solution of integral
Faddeev equation (\ref{pvg}) to the solution of the system of
linear algebraic equations (\ref{m_pvg}) and define simple
procedures and formulas for the calculation of the kernel matrix
$\mathbb K = {\mathbb P} {\mathbb V}_1 {\mathbb G}_1 $. In such an
approach, we avoided all the difficulties of solving the integral
equation ($\ref{pvg} $), which are met in the standard approach,
but the prize paid for this is a high dimension of the resulting
system of equations. This high dimension is the only  problem in
the practical solution of the  matrix analogue for the Faddeev
equation.

In fact, we found \cite{KPR_Ann} that quite satisfactory results
can be obtained with a basis size along one Jacobi momentum $N\sim
\bar{N}\sim 100-150$. It means that in the simplest one-channel
case (e.g. for $s$-wave three-boson or spin-quartet $s$-wave $nd$
scattering) one gets a kernel matrix with dimension $M=N\times
\bar{N}\sim 10000 - 20000$. However, in  case of realistic $3N$
scattering it is necessary to include at least up to 62
spin-angular channels and dimension of the matrix increases up to
$5\cdot 10^5-10^6$.  The high dimension of the algebraic system
leads to two serious problems: the impossibility to place the
whole kernel matrix into RAM and the impossibility to get the
numerical solution for a reasonable time, even using a
supercomputer.

The second obstacle can be easily circumvented.
  Indeed, to find
the elastic and breakup amplitudes one needs only on-shell matrix
elements of the transition operator. Each of these elements can be
found  by means of a simple iteration procedure (without complete
solving the matrix equation (\ref{m_pvg})) with subsequent
summation of the iterations via the well-known Pade-approximant
technique.

The first problem means that one has to store  the whole kernel
matrix in the external memory. However when using it  the
iterative process becomes very inefficient, since most of the
processing time is spent for reading data from the external memory, while
the processor is idle. Nevertheless the specific matrix structure
of the kernel in the eq.~(\ref{m_pvg}) makes it possible to
overcome this difficulty and
 {\em to eliminate completely the use of an external memory}.
 Indeed, the matrix  kernel $\mathbb K$ for
equation~(\ref{m_pvg}) can be written as a product of four matrices,
which have the specific structure:
\begin{equation}
\label{kernel} \mathbb K=\mathbb P {\mathbb V}_1  \mathbb G_1 \equiv
\mathbb O \mathbb P^0 \tilde{\mathbb V}_1 \mathbb G_1,
\end{equation}
where $\tilde{\mathbb V}_1 =  \mathbb O^{\rm T}\mathbb V_1$. Here
$\mathbb G_1$ is a diagonal matrix, $\mathbb P^0$  is a highly sparse
permutation matrix, while  $\tilde{\mathbb V}_1$ and $\mathbb O$ are
block matrices with the block dimension $(N\times N)$.

Thus, if to store in RAM only the individual multipliers of the
matrix kernel $\mathbb K$, and to store highly sparse matrix
$\mathbb P^0 $ in a compressed form (i.e. to store only its nonzero
elements), all the data required for the iteration process can still
be placed in RAM. And although in this case three extra matrix
multiplication is added at each iteration step, a computer time
spent on iterations is reduced more than 10 times in comparison with
the procedure employing an external memory.

Thus, the overall numerical scheme for solving the
three-body scattering problem in the WP discrete formalism
 consists of the following main steps:\\
 1. Processing of the input data.\\
2. Calculation of  nonzero elements of the permutation matrix ${\mathbb P}^0$.\\
3. Calculation of the channel resolvent matrix $\mathbb{G}_1$. \\
4. Iterations of the matrix  equation (\ref{m_pvg}) and finding
its
solution by making use of the Pade-approximant technique. \\

The step 1 includes the following procedures: \\
-- a construction of two-body free WP bases,
and a calculation of matrices of the interaction potential;\\
-- a diagonalization of the pairwise subHamiltonian matrices in the
free WP basis and finding parameters for the three-body channel basis including
matrices of the rotation between free and scattering WPs;\\
-- a calculation of algebraic coefficients
$g^{\Ga\al\be,\al'\be'}_{\ga\ga'}$ from eq.~(\ref{gal})  for
recoupling between different spin-angular
channels.\\

 We found that the
runtimes for the steps 1 and 3 are practically negligible in comparison with the
total running time,  so that we shall not discuss these steps here.
The execution of the step 4
--- the solution of the matrix system by iterations --- takes about
20\% of the total time needed to solve the whole problem in one-thread
CPU computing.
Therefore, in this work we did not aim to optimize this step using
the GPU.

 The main computational efforts (in the one-core CPU realization)  are
spent on the step 3 -- the calculation of elements of the matrix
${\mathbb P}^0$. Because all of these elements are calculated with
help of the same code and fully independently from each other, the
algorithm seems very suitable for a parallelization and
implementation on multiprocessor systems, in particular on GPU.
However, since the matrix ${\mathbb P}^0$ is highly sparse, it is
necessary to use special tricks in order to reach a high
acceleration degree in GPU realization. In particular, we apply an
additional pre-selection of nonzero elements of the matrix
${\mathbb P}^0$.

It should be stressed here that steps 1 and 2 do not depend on the
incident energy. The current  energy is taken into account only at
steps 3 and 4 when one calculates the channel resolvent matrix
elements and solves the matrix equation for the scattering
amplitude. Thus when one needs scattering observables in some
wide energy region, the whole computing time will not increase
sufficiently because the most time-consuming part of the code (step 2) is
carried out only once for many energy points.

\begin{figure}[h!]
\centerline{\epsfig{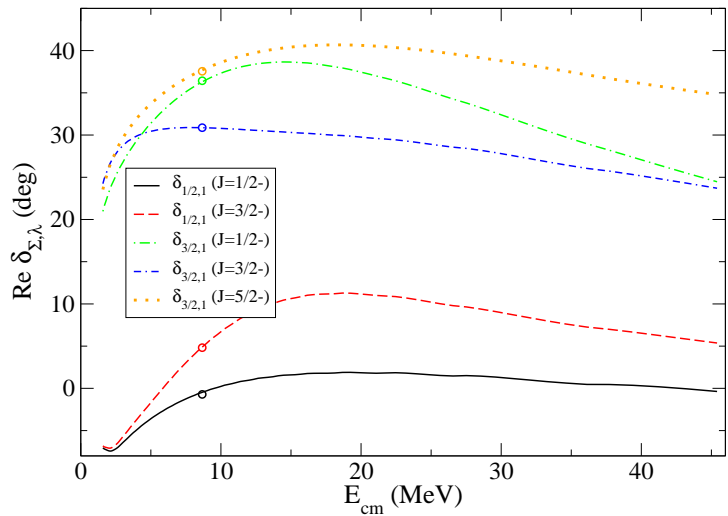}} \caption{
The $p$-wave partial phase shifts for the elastic $nd$ scattering
obtained within the WP approach (solid curves) and within the
standard Faddeev calculations
 (circles)~\cite{gloeckle}.}
 \label{phases}
\end{figure}
In  Fig.~\ref{phases}   the $p$-wave partial phase shifts
$\delta^{J\pi}_{\Sigma\lambda}$ of the elastic $nd$ scattering for
the Nijmegen I $NN$ potential \cite{nijm} both below and above a
three-body breakup threshold are shown. Here $J$, $\pi$ and
$\Sigma$ are the total angular momentum, parity and total channel spin
respectively while $\lambda$ is the neutron orbital momentum. The
calculation of the phase shifts {\em at 100 different energy
values} displayed in Fig.~\ref{phases} takes in our approach (in
CPU realization) only about twice as much time as compared with
the calculation for a single energy because for all energies  we
employ the same permutation matrix $\mathbb P$ which is calculated
only once.

In the next section we consider the specific features related to GPU
adaptation for the above numerical scheme.

%In particular, how to treat effectively a sparse matrix when doing the GPU realization.

\section{GPU acceleration in calculation of kernel matrix elements}
%%%%%%%%%%%%%%%%%%%%%%%%%%%%%%%%%%%%%%%%%%%%%%%%%%%%%%
As was noted above, the calculation of  elements of a large matrix
looks to be very suitable task for effective application of GPU
computing if these elements are calculated independently from each
other and by one code. However, there are a number of aspects
associated with the organization of the data transfer from RAM to
the GPU memory and back and also with the  GPU computation itself. These
aspects impose severe restrictions on the resulting acceleration in
GPU realization. One can introduce the  GPU acceleration $\eta$ as a
ratio of runtime for one-thread CPU computation to runtime for
the multithread GPU computation:
\begin{equation}
\label{acc} \eta=t_{\rm CPU}/t_{\rm GPU},
\end{equation}
This acceleration depends on the ratio of the actual time for the
calculation of one matrix element,  $t_0$, to the time of
transmitting the result from the GPU memory back to RAM, on the
number of GPU cores and their speed as compared to speed of CPU
core, and also on the dimension of the matrix $M$. Note that the
transition itself from a one-thread computing to multithread
computing takes some time, so that any parallelization is not
effective for matrices with low dimension. When using
the GPU, one has to take into account  that the speed of GPU cores
are usually much smaller than the CPU speed. For the efficiency of
multithread computing it is also necessary that the calculations
in all threads are finished at approximately the same time.
Otherwise a part of threads, each of which occupies a physical
core, will be idle for some time. In the case of independent
matrix elements, this condition means that the numerical code for
one element should not depend on its number, in particular, the
code must not contain conditional statements that can change the
amount of computation.

When calculating the permutation matrix ${\mathbb P}^0$ in our
algorithm, the above condition is not valid: only about 1\% of its
non-vanishing matrix elements should be really calculated using a
double numerical integration, while other 99\% of elements are equal
to zero and their  determination  requires only a few arithmetic
operations. Therefore, when one fills the whole matrix $ {\mathbb
P}^0$ (including both zero and nonzero elements) 99\% of all threads
are idle, and we will not reach any real acceleration. Thus we have
to develop at first a numerical scheme to fill effectively sparse
matrices using GPU.

\subsection{GPU acceleration in calculating elements of a sparse matrix}

In this subsection in order to check the possibility of GPU
acceleration in the calculation of the elements of a matrix with a
dimension $M$, we consider  two simple examples in which the
matrix elements
 are determined by the following formulas:\\
(a) as a sum of simple functions:
\begin{equation}
A(i,j)=\sum_{k=1}^K\left (\sin^k(u_{ij})+\cos^k(w_{ij})\right ),
{\rm or} \label{Atrig}
\end{equation}
(b) as  a sum of numerical integrals:
\begin{equation}
A(i,j)=\sum_{k=1}^K\int_{u_{ij}}^{w_{ij}}\left
(\sin^k(t)+\cos^k(t)\right )dt. \label{Agau}
\end{equation}
Here $u_{ij}$ and $w_{ij}$ are random numbers from the interval
$[0,1]$ and the parameter $K$ allows to vary the time $t_0$ for
calculation of each element in a wide range. The integrals in
eq.~(\ref{Agau}) are calculated numerically by the 48-point Gaussian
quadrature. Therefore the example (b) with numerical integration is
closer to our case of calculating the permutation matrix ${\mathbb
P}^0$ in the Faddeev kernel.

\begin{figure}[h!]
\centerline{\epsfig{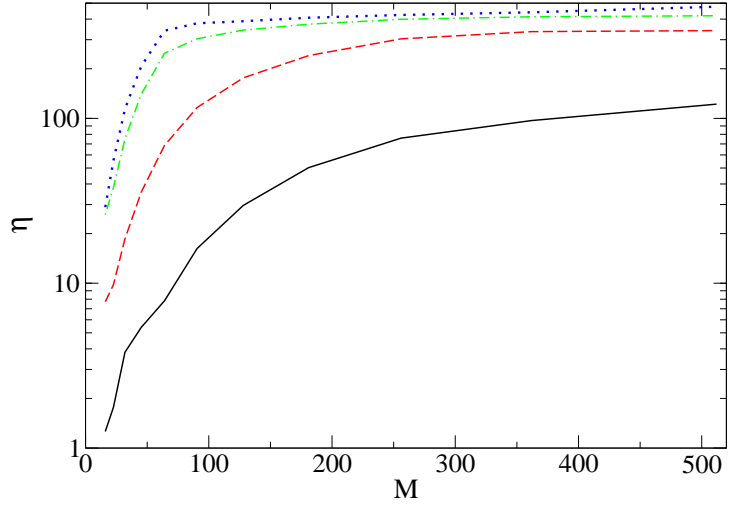}} \caption{
The dependence of GPU acceleration $\eta$ in calculation of elements
of dense matrix (\ref{Atrig}) on the matrix dimension $M$ for
different values of $t_0$: 0.0009 ms (solid curve), 0.0094 ms
(dashed curve), 0.094 ms (dot-dashed curve), 0.94 ms (dotted
curve).} \label{accel_N-trig}
\end{figure}

\begin{figure}[h!]
 \centerline{\epsfig{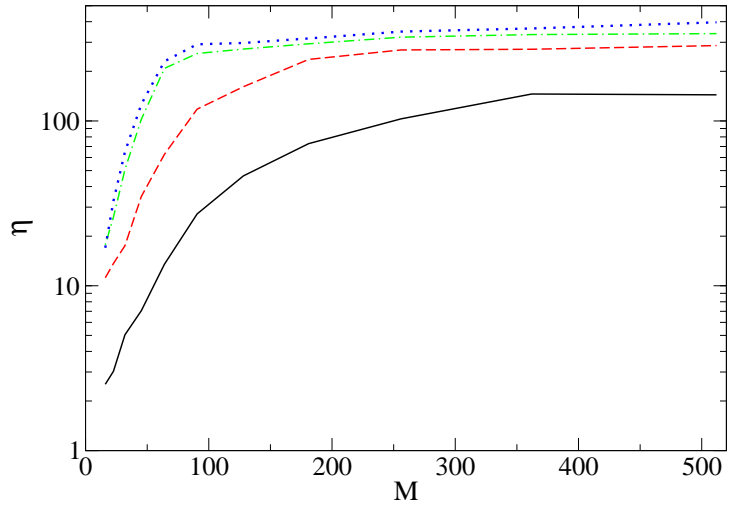}}
\caption{ The dependence of GPU acceleration $\eta$ in calculation
of elements of a dense matrix (\ref{Agau}) on the matrix dimension
$M$ for different values of $t_0$: 0.0017 ms (solid curve), 0.012 ms
(dashed curve), 0.114 ms (dot-dashed curve), 1.13 ms (dotted
curve).} \label{accel_N-gau}

\end{figure}

\begin{figure}[h!]
\centerline{\epsfig{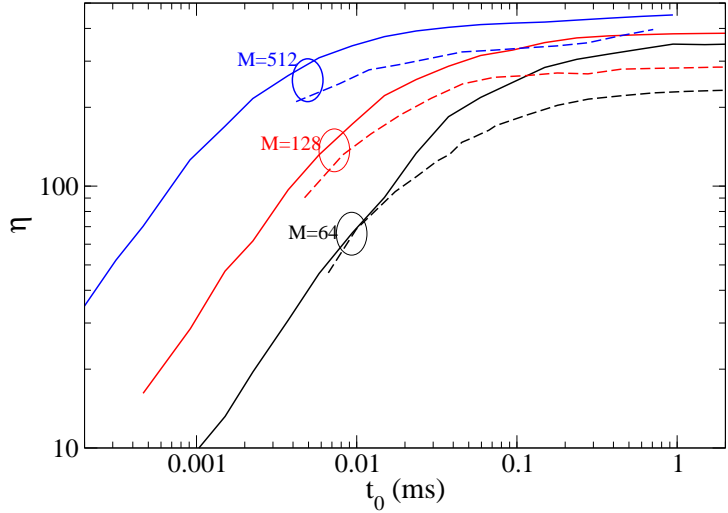}} \caption{
The dependence of GPU acceleration $\eta$ in
 calculation of elements of dense matrix on the computational time of each matrix
 element, $t_0$, for different values of matrix dimension $M$: solid
 curves
 correspond to calculation of matrix elements  using simple trigonometric
 functions  (\ref{Atrig}), dashed curve --- using numerical integrals
 (\ref{Agau}).}
\label{accel_t0}
\end{figure}

Figures \ref{accel_N-trig}, \ref{accel_N-gau} and \ref{accel_t0}
show the dependence of the GPU acceleration $\eta$  on the matrix
dimension $N$ and the calculation time for each element $t_0$ when
filling up the dense matrices defined by eqs.~(\ref{Atrig}) and
(\ref{Agau}). The GPU calculations were performed using $M^2$
threads, so that, each thread evaluates only one matrix element.


The calculations are performed on a desk PC with the processor
i7-3770K (3.50GHz) and the video card NVIDIA GTX-670. We use the
Portland Group Fortran compiler 12.10 including CUDA support and
CUDA compiler V5.5.
 As can be seen
from the figures, GPU acceleration sufficiently rises with
increasing the dimension $M$ and the computational time for one
matrix element $t_0$. The maximal acceleration that can be reached
in this model example is 400-450(!) Such high degree of
acceleration is achieved at the matrix dimension $ M\sim 200$ and
$t_0 \gtrsim 0.1$~ms. At further increase of the dimension $M$,
the degree of acceleration does not change because in this case
all the computing resources of the GPU are already exhausted. Note
that, for the example (b) with the numerical integration, the GPU
acceleration is somewhat lower than in the case of calculating
simple functions. This is due to repeated use of the some
constants (the values of the quadrature points and weights) which
should be stored in the global GPU memory.

It should also be noted that the transition to the double-precision
calculations of the matrix elements reduces greatly the maximal
possible value of GPU acceleration $\eta$.

 Consider now what efficiency of GPU computing   can be reached in the case
of a sparse matrix, when it is actually required to calculate only
part of matrix elements. We introduce the following additional
condition for the  matrix elements (\ref{Atrig}) and (\ref{Agau}):
\begin{equation} \tilde A(i,j)=\left
\{\begin{array} {cc}A(i,j), & u_{ij}\le\alpha\\ 0, &  u_{ij} >
\alpha
\end{array} \right . . \label{Acond}
\end{equation}
Since $u_{ij}$ is a random number in the interval $[0,1]$, then
one gets a sparse matrix with the degree of a sparseness
$\sim\alpha$ as a result of such filtration.  In fact, the degree
of a sparseness is the ratio of number of non-zero matrix elements
to their total number $M^2$.

\begin{figure}[h!]
\centerline{\epsfig{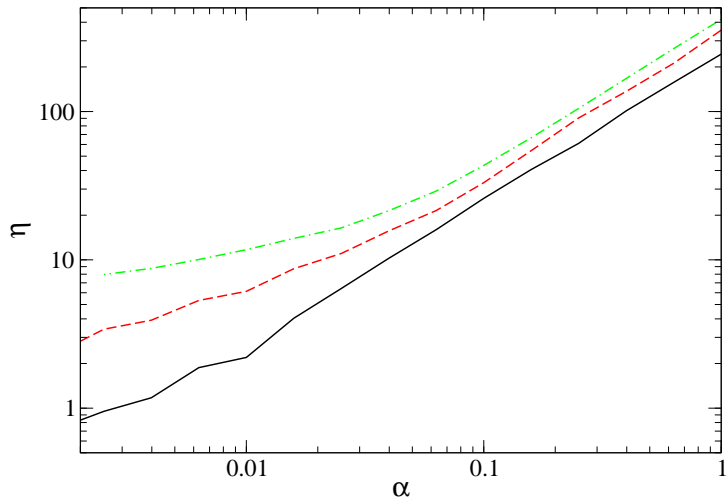}} \caption{
The dependence of GPU acceleration $\eta$ in calculation of
 elements of sparse matrix with elements (\ref{Acond}) on sparseness parameter $\alpha$:
 solid curve --- for $M=64$
 dashed curve --- for $M=128$
 solid curve --- for $M=256$.}
\label{accel_spar}
\end{figure}

Fig.~\ref{accel_spar} shows the dependence of the GPU acceleration
on the sparseness parameter $\alpha $ in filling matrices with
dimensions $M = 64$, 128 and 256. As can be seen from the Figure,
 the GPU acceleration is only about 2 (for $ M = 64 $) at a value of
 $\alpha \sim 0.01$, which corresponds to the realistic sparseness
parameter for the permutation matrix ${\mathbb P}^0$ in  the
Faddeev kernel.

Thus, to achieve  a significant GPU acceleration in calculating the
permutation matrix ${\mathbb P}^0 $, it is necessary to add one more
step to our numerical scheme discussed in the section~\ref{secIIIc}
and perform {\em a pre-selection} of nonzero
 elements of the permutation matrix.

\subsection{The  GPU algorithm for calculating the  permutation matrix  in the
case of a semi-realistic $s$-wave $NN$ interaction}
%%%%%%%%
Consider now a  calculation of the permutation matrix ${\mathbb
P}^0$  entering the Faddeev kernel. There are additional
limitations for the GPU algorithm for this case compared to
simple examples discussed in the previous subsection.

a) As already mentioned above, the most serious
limitations are a high dimension and a high sparseness of the
permutation matrix, and therefore a special packaging for this
matrix is required. Standard packaging for a matrix (we use the
packaging on the rows
--- the so called CSR format) implies, instead of storing the
matrix in a single array ${A}$ with a dimension $M\times M$,
 the presence of two linear arrays,
${B}$ and ${C}$, with dimensions $\alpha M^2$, which
store the nonzero matrix elements of ${A}$ and the respective
numbers of columns. Also the third linear array ${W}$ with
the dimension $M$ contains addresses of the last nonzero elements
(in the array ${B}$), corresponding to a given row of the
initial matrix ${A}$. With such a way of the matrix packaging
we get a gain in the memory required for storing the matrix to be
equal to $1/(2\alpha)$, i.e. about 50-fold gain for a value of the
sparseness 0.01 which is specific for the permutation matrix
$\mathbb{P}^0$ in the WP representation. So that, at the specific
matrix dimension $M\sim 5\cdot 10^5$ which is necessary for an
accurate calculation of the realistic $3N$ scattering problem, the
whole matrix occupies about 1,000 GB of RAM (with single precision),
while  the same matrix in a compressed form takes about
 20 GB RAM only. This is a quite acceptable value for a modern desktop
computer.

b) However, the permutation matrix of such a dimension, even in a
packed form, cannot be placed in the GPU memory which is usually
4-8 GB only. Therefore one needs to subdivide the whole
calculation of this matrix into some blocks using an external CPU
cycle and then employ the multithread GPU computation for each
block.

c) Another distinction of the  calculation of the elements of the
matrix ${\mathbb P}^0$ from the  simple model example discussed
above is the necessity to use a large number of constants: in
particular, the values of nodes and weights for Gaussian quadratures
for a calculation of double integrals and also (in  case of a
realistic $NN$ interaction with tensor components) algebraic
coefficients $g^{\Ga\al\be,\al'\be'}_{\ga,\ga'}$ from the
eq.~(\ref{gal}) for coupling of different spin-angular channels,
values of Legendre polynomials at the nodal points etc. All these
data are stored in the global GPU memory and because of the
relatively low access rate of each thread to the global GPU memory,
the resulted acceleration  is noticeably lower than in the case of
the above simple code which does not use a large amount of data from
the global GPU memory.

d) The necessary pre-selection of nonzero elements of the matrix
${\mathbb P}^0$ can be itself quite effectively  parallelized for
a GPU implementation. Since the runtime for checking the selection
criteria for each element is on two orders of magnitude less than
the runtime for calculating nonzero element itself, then the
degree of GPU acceleration for the stage of a pre-selection turns
out less than for the basic calculation. Nevertheless, if do not
employ the GPU at this stage, the computing time for it turns out
even larger than the GPU calculation time for all nonzero elements
(see below).

%\subsubsection{GPU acceleration in case of semi-realistic $NN$
%interaction}

After these general observations, we describe the results for the
GPU computing of the most tedious step of solving the Faddeev
equation in the WP  approach --- the computation of nonzero
elements of the permutation matrix --- in the case of a
semi-realistic Malfliet-Tjon $NN$ interaction. There is no
spin-angular coupling for this potential, so that the Faddeev
system  is reduced  to a single $s$-wave equation. The results
attained for a realistic calculation of multichannel $ nd $
scattering we leave for the next section.

When the pre-selection of nonzero matrix elements is already done
 one has the subsidiary arrays ${C}$ and ${W}$ containing information
about all nonzero elements of $\mathbb{P}^0$ that should be
calculated and the number of these nonzero elements is $M_t$. The
parallelization algorithm adapted here assumes that every matrix
element is computed by a separate thread. The allowable number of
threads $N_{\rm thr}$ is restricted by the capacity of the physical GPU
memory and is usually less than the total number of nonzero elements $M_t$.
In this case, our algorithm consists of the following steps.

1. The data used in  calculation (endpoints of momentum intervals
in variables $ p $ and $ q $, nodes and weights of Gauss
quadratures, algebraic coupling coefficients etc.) are copied to
the GPU memory.

2. The whole set  of nonzero elements of the permutation matrix is
divided into $N_b$ blocks with  $N_{\rm thr} $ elements in each block
(except the last one) and the external CPU loop is organized by the
number of such blocks. Inside the loop the following operations are
performed:

3. A part of the array ${C}$ corresponding to the current
block is copied to the GPU memory.

4. The CUDA-kernel is launched on GPU in $N_{\rm thr}$ parallel
threads each of which calculates only one element (in the case of
the $s$-wave problem) of the permutation matrix.

5. The resulted $N_{\rm thr}$ nonzero elements of the matrix are
copied from the GPU memory to the appropriate place of the total
array ${B}$.

\begin{figure}[h!]
\centerline{\epsfig{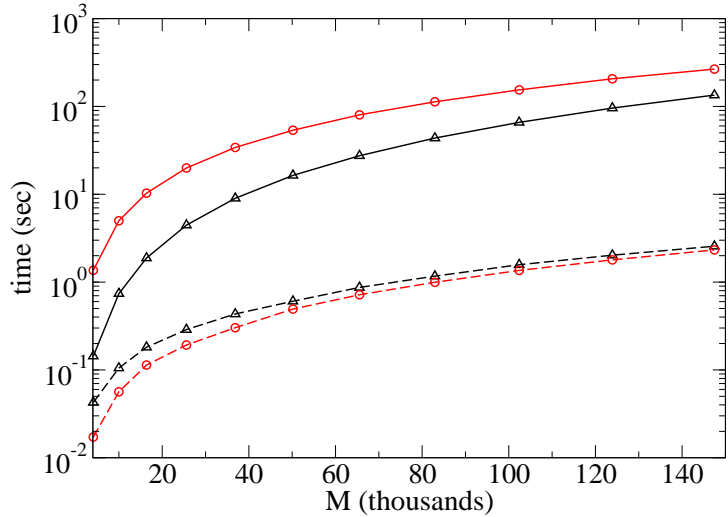}} \caption{
The CPU computing time (solid curves) and GPU computing time (dashed
curves)  for preselection of tne nonzero elements of $s$-wave
permutation matrix ${\mathbb P}^0$ (triangles) and  calculation of
these elements (circles) depending on the matrix dimension $M$.}
 \label{time_GPU-sw}
\end{figure}

Fig.~\ref{time_GPU-sw} shows the dependence of the CPU- and
GPU-computing time for the calculation of the $s$-wave permutation
matrix  upon its total dimension $M=N\times\bar{N}$ (for
$N=\bar{N}$). In our case, the GPU code was executed in 65536
threads. For the comparison, we display on this Figure also the CPU
and GPU time which are necessary for a pre-selection of nonzero
matrix elements. It is clear from the Figure that  one needs to use
GPU computing not only for the calculation of nonzero elements (that
takes most of the time in one-thread CPU computing), but also for
the pre-selection of nonzero matrix elements to achieve a high
degree of the acceleration.

\begin{figure}[h!]
\centerline{ \epsfig{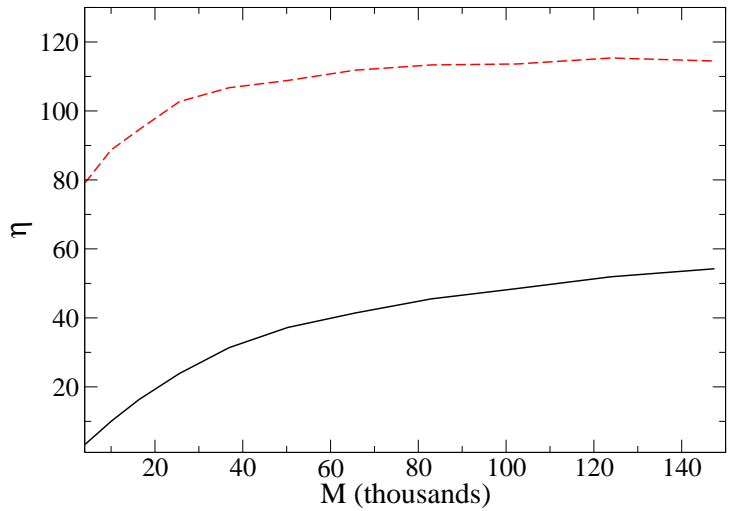}} \caption{
The dependence of the  GPU acceleration
 $\eta$ on the matrix dimension  $M$ for a calculation of
the permutation matrix  (dashed curve) and for a complete solution
of the scattering problem (solid curve) in the case of $s$-wave
$NN$ interaction.} \label{accel_GPU-sw}
\end{figure}

In  Fig.~\ref{accel_GPU-sw}, we present the GPU acceleration
$\eta$ for calculating the $s$-wave permutation matrix  and for a
complete solution of $s$-wave $nd$ elastic scattering problem  on
dimension  $M$ of the matrix equation. It is evident that the
runtime for the nonzero elements of the matrix ${\mathbb P}^0$
(which takes the main part of the CPU computing time) is reduced
by more than 100 times. The total acceleration in calculating the
$s$-wave partial phase shifts reaches 50.
 Finally, the total three-body calculation takes only 7 sec. on
  an ordinary PC with GPU.

\section{GPU optimization for a realistic $3N$ scattering problem}
%%%%%%%%%%%%%%%%%%%%%%%%%%%%%%%%
\subsection{GPU-acceleration for a realistic $nd$ scattering amplitude}
%%%%%%%%%%%%%%%%%%%%%%%%%%%%%%%%%%%%%%%%%%%%%%
We now turn to the case of a realistic three-nucleon scattering
problem with the Nijmegen I $NN$ potential \cite{nijm} and the
calculation for the  elastic $nd$ scattering cross section.

Unlike the simple $s$-wave case discussed above, now we have many
coupled spin-angular channels (up to 62 channels if the total
angular momentum in $NN$ pair is restricted as $j\le 3$). In this
case, the calculation of each element of the permutation matrix
${\mathbb P}^0$ comprises the calculation of  several tens of
double numerical integrals containing the Legendre polynomials.
Each matrix element is equal to the sum of such double integrals
and the sum includes a large set of algebraic coupling
coefficients $g^{\Ga\al\be,\al'\be'}_{\ga,\ga'}$ for the spin-angular
channels as in eq.~(\ref{gal}).

Now the GPU-optimized algorithm for the  permutation matrix is
somewhat different: because each  calculated double integral is
used to compute several matrix elements, then each thread now
calculates all the matrix elements corresponding to one pair of
momentum cells $\{\MD_{ij},\MD_{i'j'}\}$. These matrix elements
belong to
 different rows of the complete permutation matrix. So that,
after the GPU computing for each block of the permutation matrix
it is necessary to rearrange and repack (in the single-thread CPU
execution) the calculated set of the matrix elements into the
arrays ${B}$, ${C}$ and ${W}$, representing
the complete matrix $\mathbb{P}^0$ in CSR format. All the above
leads to the fact that the GPU acceleration in calculation of the
permutation matrix in a realistic case when the $NN$ interaction
has a tensor component  turns out significantly less than for the
$s$-wave case.

\begin{figure}[h!]
 \centerline{ \epsfig{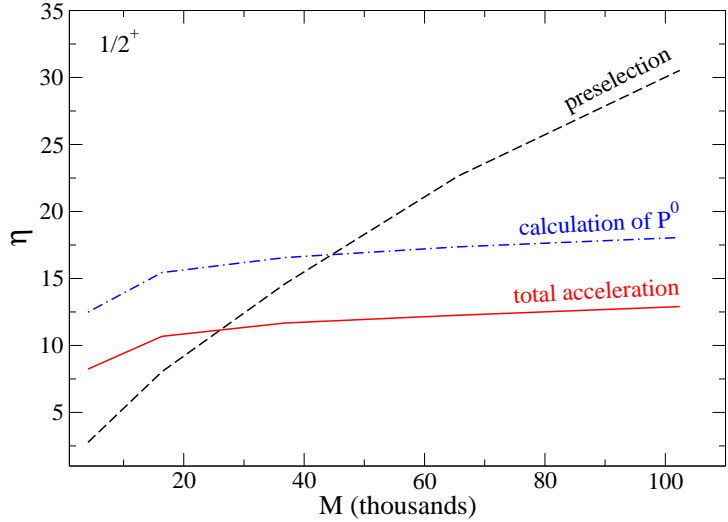}}
\caption{  The dependence of the GPU acceleration
 $\eta$ on dimension of the basis $M=N\times \bar{N}$ (for the case
$N=\bar{N}$) for the realistic $nd$ scattering problem at
$J=\frac12^+$: dashed line shows the acceleration for the
preselection of nonzero elements in the permutation matrix
 ${\mathbb P}^0$),
 dot-dashed line -- for the calculation of these elements,
 solid line -- the  acceleration for the complete solution.
 \label{time_GPU_Nijm}}
\end{figure}

 Fig.~\ref{time_GPU_Nijm} demonstrates the GPU acceleration
 $\eta$ versus the basis dimension $M=N\times\bar{N}$ in the solution
 of 18-channel Faddeev equation for the partial
$nd$ elastic amplitude with total angular momentum $J=\frac12^+$ (solid line).
The dashed and dot-dashed lines show the GPU acceleration for stage of
 pre-selection of nonzero elements for the permutation matrix ${\mathbb P}^0$ and
 for calculating of these elements, respectively.

 From these results, it is evident that the acceleration in the calculation of the
coupled-channel permutation matrix is about 15 that is
considerably less in comparison with the above one-channel
$s$-wave case. Nevertheless, the passing from CPU- to
GPU-realization {\em on the same PC} allows to obtain a quite
impressive  acceleration about 10 in the solution of the
18-channel scattering problem.

\begin{figure}[h!]
 \centerline{ \epsfig{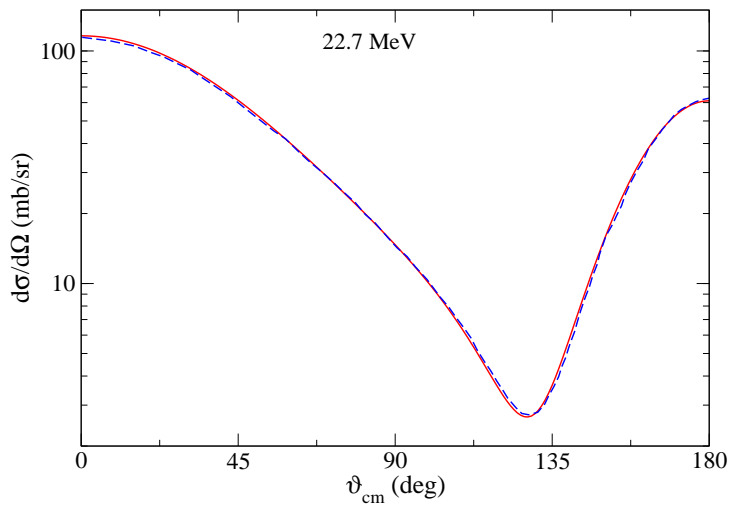}}
\caption{  The differential cross section of elastic $nd$ scattering
at energy 22.7 MeV calculated with the Nijmegen I $NN$ potential in
wave-packet formalism with using GPU computing (solid curve) in
comparison with the results from the ref.~\cite{gloeckle} (dashed
curve).}
 \label{diff_cs}
\end{figure}

In realistic calculation of the observables for elastic $nd$
scattering, it is necessary to include up to 62 spin-orbital
channels.  For the current numerical scheme, the efficiency of GPU
optimization decreases with increasing number of channels. As an
example, we present the results of the complete calculation for
elastic $nd$ scattering with the Nijmegen I $NN$ potential at
energy 22.7 MeV. In Fig.~\ref{diff_cs} as an illustration of an
accuracy  of our approach, we display the differential cross
section  in comparison with the results of the conventional
approach~\cite{gloeckle}.

%However, it should be emphasized that
%acceleration for complete solution which can be achieved by using GPU depends crucially
%on the method used, the numerical scheme and parallelization way.

The complete calculation, including 62 spin-orbital channels and all
states with total angular momentum up to $J_{\rm max}=17/2$ took
about 30 min on our desk PC. The runtimes for separate steps are given in
Table 1.

\begin{table}[h!]
\centering \caption{Runtime (in sec) for separate steps of complete
solutions of $nd$ scattering problem}
\begin{tabular}{llcc}
&Step & CPU time & GPU time\\ \hline
1.  & Processing input data   & 30        & 30\\
2a.& Pre-selection & 12 & 1.9 \\
2b.& Calculation of nonzero elements & 4558 & 524\\
4. &Iterations and Pade summation & 1253&1250\\ \hline
   &Total time & 5852 & 1803 \\ \hline
\end{tabular}
\end{table}

As seen from the Table,  the time of calculation of the
permutation matrix elements  (steps 2a and 2b) is shorten in ca.
8.7 times as a result of the GPU optimization.  However, the major
part of a computational time
 is now spent not on calculating
the permutation matrix but  on  the subsequent iterations of the
resulting matrix equation, i.e. on multiplication of a kernel matrix
by a column of current solution. The iteration time takes about 69\% of total solution time.
 So that, the total acceleration in this multichannel case is only 3.2.

 It should be stressed that the current numerical scheme can be further optimized.
 Each  iteration here includes four
matrix multiplications: one multiplication by a diagonal matrix
$\mathbb G_1$, two multiplications by block matrices $\mathbb O$ and
$\tilde {\mathbb V}_1$ and one multiplication by sparse matrix
$\mathbb P^0$, and most of the time in the iteration process
 takes multiplication of a sparse matrix by a (dense) vector.
It is clear that the algorithm for the iteration can also be
parallelized and implemented on the GPU. In this paper, we did not
addressed this task and focused mainly on the GPU optimization for
the calculation  the integral kernel of Faddeev equation only.
However, for a multiplication of a sparse matrix to a column there
are standard procedures, including those implemented on GPU. So
that, if to apply the GPU optimization to the iteration step the
runtime of complete solution can be reduced further by 2-3 times.

It is also clear that
employment of more powerful specialized graphics processors would
lead even to a considerably greater acceleration of the
calculations.

\subsection{Further development}
%%%%%%%%%%%%%%%%%%%%%%%%%%%%%%%%%%%%%%%%%%%%%%%%%%
 It looks evident that the described GPU approach will be
effective also in the solution of integral equations describing
the scattering in systems of four and a larger number of particles
(Faddeev--Yakubovsky equations). The main difference in these more
complicated problems from  the three-body scattering problem
considered here is increasing number of channels to be included
and also rising of the dimension for integrals those define the
kernel matrix elements. As the result, the matrix dimension $M$
and the computational time of each matrix element $t_0$ will
increase. However, a degree of sparseness for the permutation
matrices and scheme for calculation of kernel matrix elements will
remain the same as in a three-body case. So that, these two
factors, i.e. growth of $M$ and $t_0$, according to our results,
will provide even greater GPU acceleration than in a three-body
case.

However, when the matrix size $M$ will reach a certain limit, no
package will be able to place all nonzero elements in RAM of a
computer. In such a case, it should be chosen another strategy: one
divides  the channel space onto two parts: the  major and minor
channels according to their influence to the resulted amplitude. The
minor channels would give only a small correction contribution to
the solution resulting from the subspace of the major channels.
Then, using the convenient projection formalism (such as the known
Feshbach formalism), one can account for the minor-channel
contribution in a matrix kernel defined in the subspace of the major
channels as some additional effective interaction containing the
total resolvent in the minor-channel subspace. We have shown
previously \cite{KPR_Ann, Moro} that the basis dimension for the
minor  channels can be considerably reduced (for a particular
problem, it can be reduced in 10 times \cite{Moro}) without loss in
an accuracy of a complete solution.

We hope that such a combined approach together with the
multithread GPU computing will lead to the greater progress in the
exact numerical solution of quantum few-body scattering problems
when using a desktop PC.

\section{Conclusion}

In the present paper we have checked the applicability of the
GPU-computing technique in few-body scattering calculations. For
this purpose we have used the wave-packet continuum discretization
approach in which
 a continuous spectrum of the
Hamiltonian is approximated by a discrete spectrum of the $L_2$
normalizable wave-packet states.  If to project out all the wave
functions and scattering operators onto such a discrete basis we
arrive at simple linear matrix equation with non-singular matrix
elements instead of the complicated multi-dimensional singular
equations in the initial formulation of few-body scattering problem.
Moreover, the matrix elements of all the constituents of this
equation are calculated independently which make the numerical
scheme to be highly parallelized.

The prize for this matrix reduction is a high dimension for the
matrix kernel. In the case of fully realistic problem the dimension
of the kernel matrix turns out so high that such a matrix cannot be
placed into RAM of a desktop PC. In addition the calculation of all
kernel matrix elements requires a huge computing time in sequential
one-thread execution. However, we have developed efficient
algorithms of parallelization, which allows to perform basic
calculations in the multithread GPU execution and reach a noticeable
acceleration of calculations.

 It is shown that the acceleration
obtained due to GPU-realization depends on the dimension of the
basis used and the complexity of the problem. So, in the three-body
problem of the elastic $nd$ scattering   with a semi-realistic
$s$-wave $NN$ interaction, we obtained 50-fold acceleration for the
whole solution while for a separate part of the numerical scheme
(most time consuming on CPU) the acceleration achieves more than 100
times. In a case of the fully realistic $NN$ interaction for the
$nd$ scattering (including up to 62 spin-orbit channels),  the
acceleration for the permutation matrix calculation is about 8.7
times.  A full calculation of the differential cross section  is
accelerated in this case by 3.2 times. However, the numerical scheme
allows a subsequent optimization that will be done in our further
investigations. Nevertheless, the present study has shown that the
implementation of GPU calculations in few-body scattering problems
is very perspective at all and opens new possibilities for a wide
group of researches.

It should be stressed, the developed GPU accelerated discrete
approach to  solution of  quantum scattering problems can be
transferred without major changes to other areas of quantum physics,
as well as to  a number of important areas of classical physics
involving solution of multidimensional problems for continuous media
studies.

 {\bf Acknowledgments}  This work has been supported
 partially by the Russian Foundation for Basic Research,
 grant No.  13-02-00399.

\end{document}